\documentclass[reprint,amsmath,amssymb,showpacs,aps,prb,superscriptaddress]{revtex4-1}

\usepackage{graphicx}
\usepackage{dcolumn}
\usepackage{bm}

\begin{document}

\title{Spin excitations in an all-organic double quantum dot molecule}

\author{Max Koole}
\affiliation{Kavli Institute of Nanoscience, Delft University of Technology, Lorentzweg 1, 2628 CJ, Delft, Netherlands}
\author{Jan C. Hummelen}
\affiliation{Stratingh Institute for Chemistry and Zernike Institute for Advanced Materials, University of Groningen, Nijenborgh 4, 9747 AG, Groningen, Netherlands}
\author{Herre S.J. van der Zant}
\email{H.S.J.vanderZant@tudelft.nl}
\affiliation{Kavli Institute of Nanoscience, Delft University of Technology, Lorentzweg 1, 2628 CJ, Delft, Netherlands}

\date{\today}

\begin{abstract}
We realize a strongly coupled double quantum dot in a single all-organic molecule by introducing a non-conjugated bridge in between two identical conjugated moieties. Spin-1/2 Kondo and Kondo enhanced low-energy excitations for respectively the odd and even electron occupation are observed in off-resonant transport. The ground state in the even occupation can be the singlet or the triplet state varying between samples. This observation suggests that both anti-ferromagnetic and ferromagnetic interactions between spins are of the same order of magnitude.
\end{abstract}


\maketitle

\section{Introduction}

Electron transport through double quantum dots made it possible to investigate new avenues of physics, such as the study of Kondo effects beyond the single-impurity Anderson model\cite{Jeong2001} and the realization of qubits\cite{Hanson2007}. Experimental systems encompass coupled semiconductor\cite{Wiel2002a}, graphene\cite{Molitor2009} and carbon nanotube\cite{Mason2004} quantum dots and quantum dots built from atom pairs\cite{Roche2012}. For single molecules, electron transport through an organic double quantum dot (DQD) molecule has not been studied to the same extend despite the fact that the energy scales of its electronic states are generally larger than in its counterparts, therefore relaxing the requirements for millikelvin temperatures for operation. Furthermore, the use of organic atoms like carbon and oxygen is expected to result in small spin-orbit coupling. This suggest that an all organic DQD molecule can potentially be a model system for the study of spin interactions that can be chemically tuned.\cite{Mujica-Martinez2013}

\begin{figure}[b]
	\includegraphics[width=0.79\columnwidth]{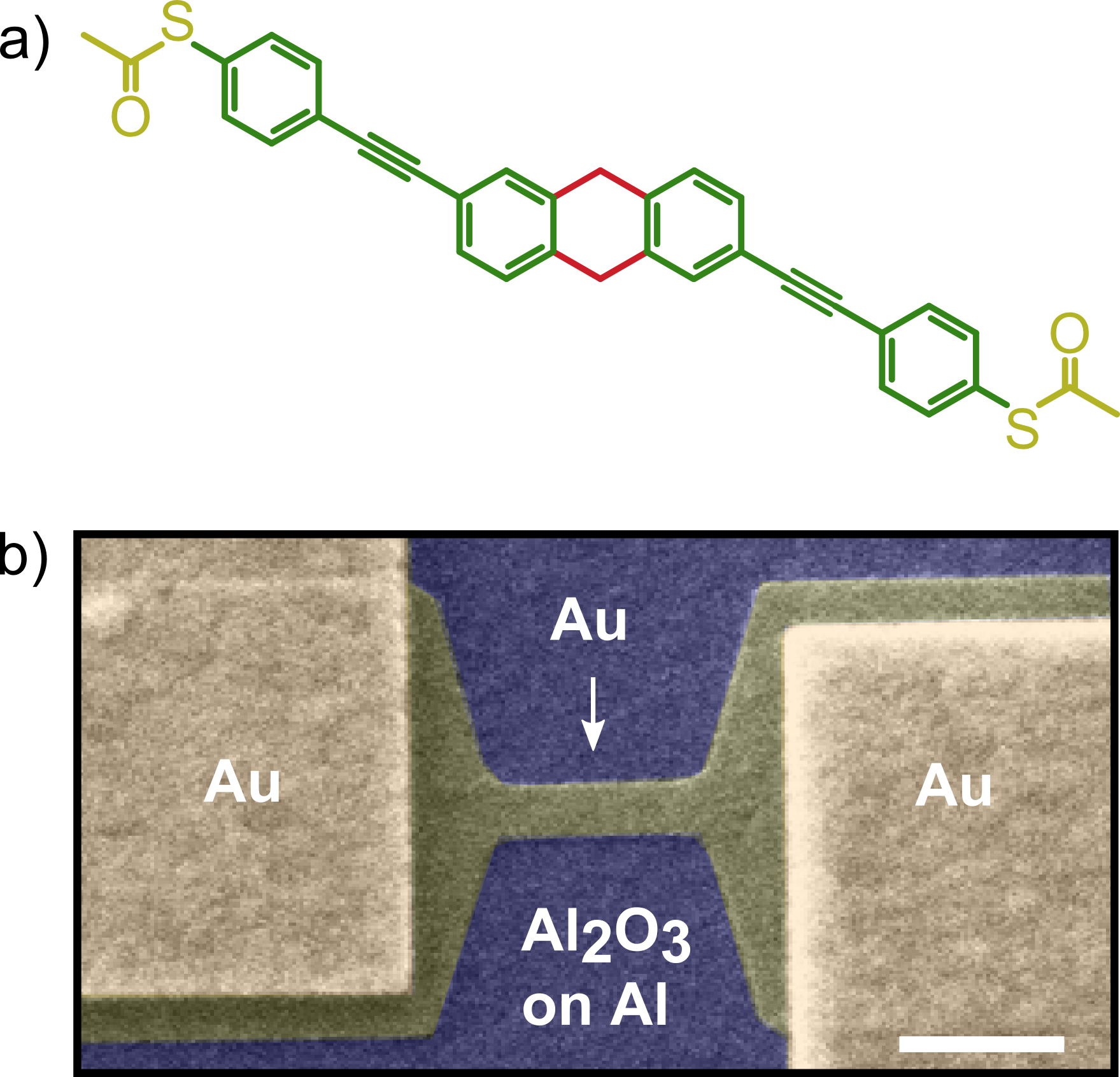}
	\caption{\label{figure:1} Molecule and setup. a) 9,10-dihydroanthracene (AH) with thiolated spacers. The dark green parts are the conjugated halves and the red colored part in the middle is the non-conjugated bridge. The yellowish colored parts at the end of the molecule are the anchoring groups. b) False-colored scanning electron microscopy image of a junction before electromigration. The purple area is the aluminum oxide surrounding the aluminum gate electrode underneath it. The darker green area is the 10 nm thick gold nano-wire. Lighter green areas correspond to the thicker leads, which are connected to wire bond pads. The white bar at the bottom has a length of 250 nm. }
\end{figure}

To study transport in an organic DQD molecule, we use a 9,10-dihydroanthracene core with acetyl-protected sulfur anchoring groups connected by spacers\cite{Fracasso2011}(see figure~\ref{figure:1}a; named AH from now on). The conjugation of AH is broken due to the sp\textsuperscript{3} hybridization of the two carbon atoms in the center of the 9,10-dihydroanthracene core. This results in AH being made up of two conjugated halves (dark green in figure~\ref{figure:1}a) connected by a non-conjugated bridge (red in figure~\ref{figure:1}a). This peculiar electronic structure results in a decreased conductance\cite{Kaliginedi2012} and a higher energy UV-VIS absorption threshold\cite{Valkenier2011} compared to the fully conjugated version of the molecule. The electronic structure of AH also results in negative differential conductance in charge transport\cite{Perrin2014}, which is explained by a two-level model formed by the broken conjugation. Past research has thus shown that the broken conjugation influences the electronic structure and transport, suggesting that it may be regarded as a molecular DQD. In this paper, we explore this possibility by using low-temperature transport spectroscopy measurements in a three-terminal configuration.

\section{Measurements}

To form a gold-molecule-gold junction, we employ three-terminal electromigration junctions (figure~\ref{figure:1}b). A silicon/silicon oxide substrate is used, on top of which an aluminum gate is deposited by e-beam evaporation. Under a pure oxygen atmosphere, the aluminum is oxidized to create an electrically insulating layer surrounding the gate terminal. On top of the gate a 12 nm thick, 100 nm wide gold nano-wire is subsequently deposited, which is connected with 100 nm thick Au leads to bond pads. To create the nano-gap the thin wire is controllably electromigrated \cite{Strachan2005} at room-temperature in a solution of dichloromethane with 0.5 mM AH to a resistance of the order of 5 K$\Omega$. After this the wire is let to self-brake \cite{ONeill2007} to avoid the formation of spurious gold nano-particles when opening a gap in the wire to form the junction. When junction resistances are of the order of 1 M$\Omega$, the sample chamber is pumped and cooled to cryogenic temperatures. The three-terminal geometry of these junctions makes it possible to measure the current ($I$) as a function of bias ($V$) and gate ($V_g$) voltages. Furthermore, with the help of a 1K-pot, heater resistor and a superconducting magnet, temperature and magnetic-field dependent measurements can be performed.

\begin{figure}
	\includegraphics[width=0.79\columnwidth]{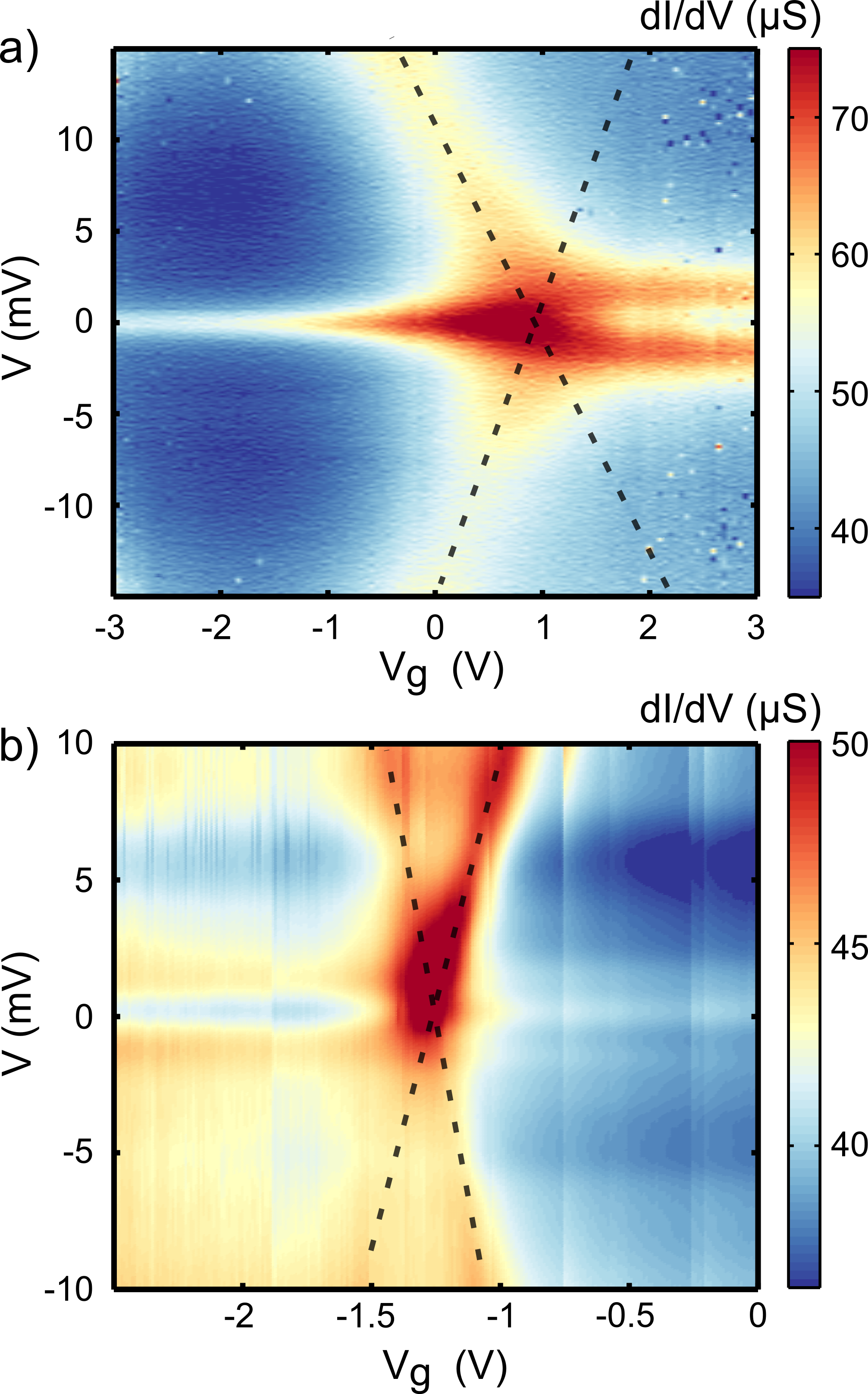}
	\caption{\label{figure:2}Differential conductance map as a function of bias ($V$) and gate ($V_g$) voltage of electromigration junction S1 (a) ($T = 2.0$~K, $B = 0$~T) and junction S2 (b) ($T = 2.0$~K, $B = 0$~T). Dashed lines show the SET edges. In a) a zero-bias peak is present in the left charge state and an excitation at a few meV in the right charge state. In b) a weak zero-bias peak (white line, see figure P7 in the supporting information for a linecut)\cite{suppinfo} is present in the right charge state and an excitation of a few meV in the left one.}
\end{figure}

Figure~\ref{figure:2}a shows a differential conductance map of a junction (S1) prepared as described above. A single-electron tunneling (SET) regime is present around $V_g = 1$~V, marked by the dashed lines. The SET edges of the left charge state are visible; the edges of the right charge state are, however, suppressed. We first focus on the left charge state in figure~\ref{figure:2}a. It shows a pronounced gate-dependent zero-bias peak. The peak splits as a function of magnetic field and its height decays exponentially as a function of temperature (see supporting information)\cite{suppinfo}. This behavior suggests spin-1/2 Kondo physics to be the origin of the zero-bias peak; the g-factor determined from the magnetic field dependent data is $g \sim 2.6$, which matches with values observed for single molecules\cite{Liu2015}. The presence of a spin-1/2 Kondo peak indicates an uneven electron occupation for the left charge state.

\begin{figure*} 
	\includegraphics{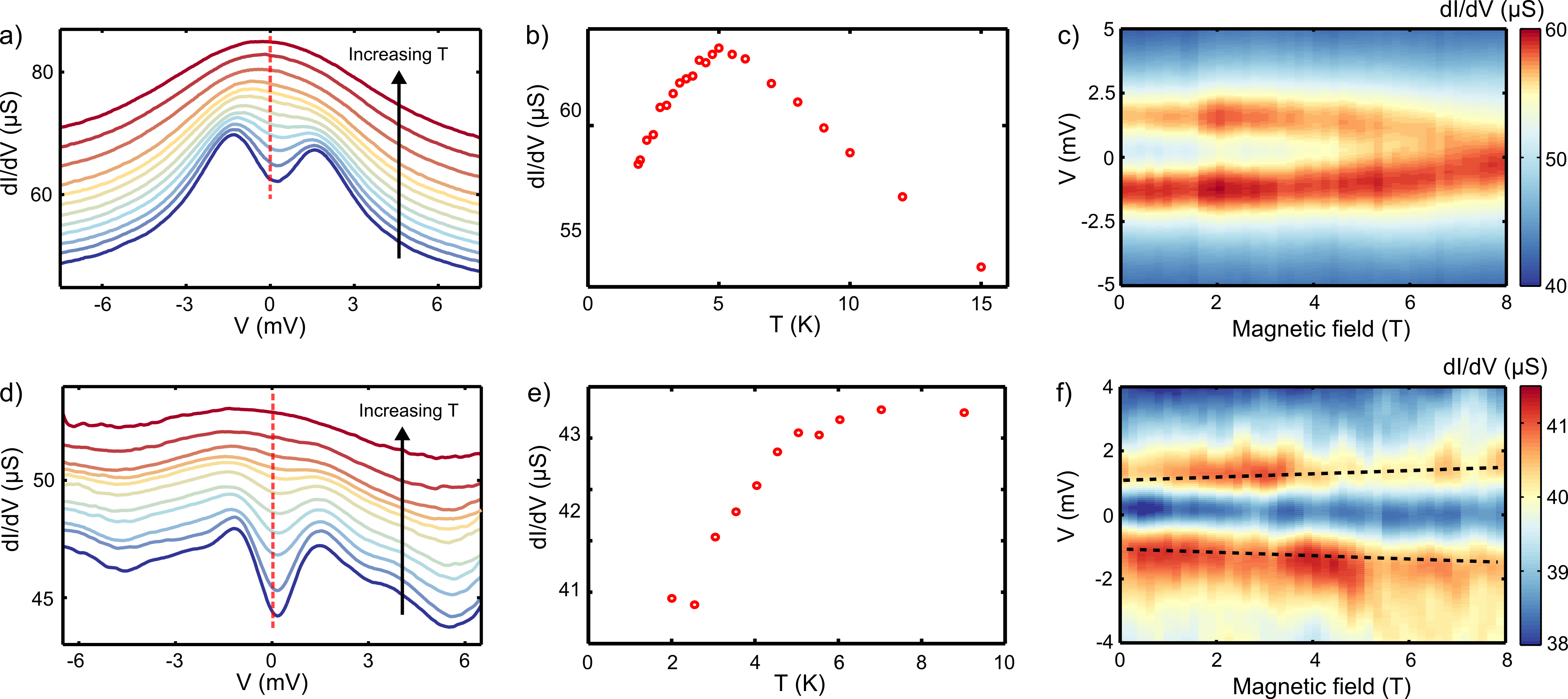}
	\caption{\label{figure:3} Temperature and magnetic field dependence of low-bias features in the even charge state for sample S1 (a,b,c) and S2 (d,e,f). a) Temperature dependence of the differential conductance at $V_g = 3$~V. The individual curves are taken at a temperature of 1.9, 2.5, 3.0, 3.5, 4.0, 4.5, 5.0, 5.5, 6.0, 7.0, 8.0 and 9.0 K and are off-set from each other. The red dashed line is the zero-bias conductance plotted in panel b. b) Zero-bias conductance as a function of temperature. c) Magnetic-field dependence of the low-bias feature in the right charge state ($V_g = 3$~V, $T = 2$~K). d,e,f) show the same measurements as a,b,c) with curves taken at 1.95, 2.5, 3.0, 3.5, 4.0, 4.5, 5.0, 5.5, 6.0, 7.0 and 9.0 K for sample S2. In d,e) $V_g = -1.9$~V and in f) $T = 2$~K, $V_g = -1.9$~V.}
\end{figure*}

When the gate voltage in figure~\ref{figure:2}a is increased past $V_g= 1$~V, an electron is added to the molecule. Instead of a zero-bias peak, a broader peak appears with a suppression around zero bias. Figure~\ref{figure:3}a shows the temperature dependence of this feature. With increasing temperature the suppression is lifted and a single broad peak remains above $T = 5$~K. Increasing the temperature even further results in a decrease in the height of the broad peak. This is seen more clearly in figure~\ref{figure:3}b, which shows the zero-bias conductance at $V_g = 3$~V as a function of temperature. This temperature dependence resembles a two-stage Kondo process\cite{Wiel2002}, which can be caused by the full screening of a triplet\cite{Pustilnik2001} or a nearly degenerate singlet/triplet state\cite{Hofstetter2001}. These scenario's are further supported by the fact that in this charge state there should be an even number of electrons on the molecule, considering the odd occupation of the adjacent charge state to the left. 

The magnetic field dependence in figure~\ref{figure:3}c shows that the suppression is lifted as a function of increasing magnetic field. This observation matches a singlet ground state ($S=0$), because in case of a triplet ground state the excited states would move up in energy when increasing the magnetic field. Taking all information together, we assign the feature in the right charge state to two spins that interact anti-ferro magnetically; the ground state is a singlet and the peak at finite bias in $dI/dV$ is caused by the singlet to triplet excitation. Using a Heisenberg hamiltonian $\mathbf{H} = -J \mathbf{S_1 \cdot S_2}$, the exchange coupling ($J$) can be determined from the figure, $J = -0.7$~meV. It is also interesting to note that when the suppression is lifted (when the singlet and triplet state are brought into degeneracy), the peak full width half maximum (FWHM) is larger than the spin-1/2 Kondo FWHM in the adjacent charge state at the same temperature. This matches other observations where a Kondo resonance caused by a singlet-triplet degeneracy has a larger FWHM than the one of the spin-1/2 Kondo\cite{Sasaki2000}.  

Figure~\ref{figure:2}b shows another sample (S2) which has two charge states separated by a SET regime around $V_g = -1.3~V$ (marked by the dashed lines). However, in this sample a zero-bias peak is present in the right charge state. Temperature and magnetic field dependent data point again to a spin-1/2 Kondo origin and thus suggest an uneven occupation of the molecule in this charge state (see supporting information)\cite{suppinfo}. Going to the left charge state by removing an electron from the molecule results in the appearance of a low-bias excitation. Figure~\ref{figure:3}d and e show the temperature dependence and figure~\ref{figure:3}f shows the magnetic-field dependence of this low-bias feature. A strong dependence on temperature is seen, suggesting Kondo correlations to play a role. Furthermore, the magnetic field dependence shows that the excitation energy shifts outwards to higher energies with increasing magnetic field. The presence of an even number of electrons on the molecule and the magnetic field and temperature dependences suggest a triplet ($S=1$) ground state with a Kondo enhanced excitation to the singlet state (the exchange coupling $J$ is $+0.5$~meV and ferro-magnetic)\cite{Sasaki2000}. Sample S2 therefore has two differences compared to sample S1; the ground state in the even charge state is a triplet instead of a singlet state and the spin-1/2 Kondo effect is found to the right of an even charge state with low-bias features instead of to the left of such a state. 

The pattern of spin-1/2 Kondo physics in one charge state and a low-bias feature due to the excitation between the singlet and triplet in a neighboring charge state has been seen in eight samples (see supporting information)\cite{suppinfo}. Table~\ref{tab:parameters} shows the addition energy ($U_{add}$), the level broadening of the SET edge ($\Gamma$), the magnitude and sign of the exchange coupling ($J$) of the eight samples (for an overview of all parameters used in this text see supporting information)\cite{suppinfo}. It can be seen that the addition energies are relatively small compared to those of molecules in gas phase. This may in part be due to the small energy difference between the HOMO and HOMO-1 (see discussion), but re-normalization of the energy levels (image charge effects associated with the gate and electrodes) may also play a role\cite{Kaasbjerg2008,Perrin2013,Kubatkin2003}. The level broadening is smaller than the addition energy, but still of the same order explaining why Kondo correlations can be seen in the measurements. The magnitude of the coupling between the spins ($|J|$) ranges from approximately degenerate up to 5 meV. The sign of $J$ has been determined for 2 samples (S1 and S2, anti-ferromagnetic and ferro-magnetic respectively). Furthermore for sample S6 and S7 the singlet and triplet states are approximately degenerate (see figure P7 in the supporting information)\cite{suppinfo}. For 4 samples magnetic field data have not been recorded so the sign of $J$ could not be determined.  

\begin{table}[b]
\caption{\label{tab:parameters} Parameters extracted from the differential conductance maps (see also supporting information)\cite{suppinfo}. Some values could not be determined; in those cases a upper or lower limit is given. $U_{add}$ is the addition energy of the specified charge state with an estimated error of $\pm 5$~meV, $\Gamma$ is the level broadening ($\pm 2$~meV) and $|J|$ the magnitude of the exchange coupling ($\pm 0.2$~meV). The last column displays the character of the coupling between the two spins, which can be anti-ferromagnetic (AF), ferromagnetic (F), degenerate (degen.) or not determined (not det.) due to the absence of magnetic field dependent data.}
\begin{ruledtabular}
\begin{tabular}{ccccc}
& $U_{add}$ (meV) & $\Gamma$ (meV) &$|J|$ (meV) & coupling \\ 
\hline
S1 & 49 (N=-1) & 11 & 0.7 & AF\\
S2 & $>$63 (N=-2) & 5.5 & 0.5 & F\\
S3 & 50 (N=-2) & 9 & 1.4 & not det. \\
S4 & 73 (N=-2) & 5 & 3 & not det. \\
S5 & 37 (N=-2)& 8 & 1.2 & not det. \\
S6 & 58 (N=-2) & 10 & $<$0.2 & degen. \\
S7 & $>$53 (N=-1) & 18 & $<$0.5 & degen. \\
S8 & $>$50 (N=-2) & 40 & 3.5 & not det. \\
\end{tabular}
\end{ruledtabular}
\end{table}

\section{Discussion}

\begin{figure} 
	\includegraphics[width=0.79\columnwidth]{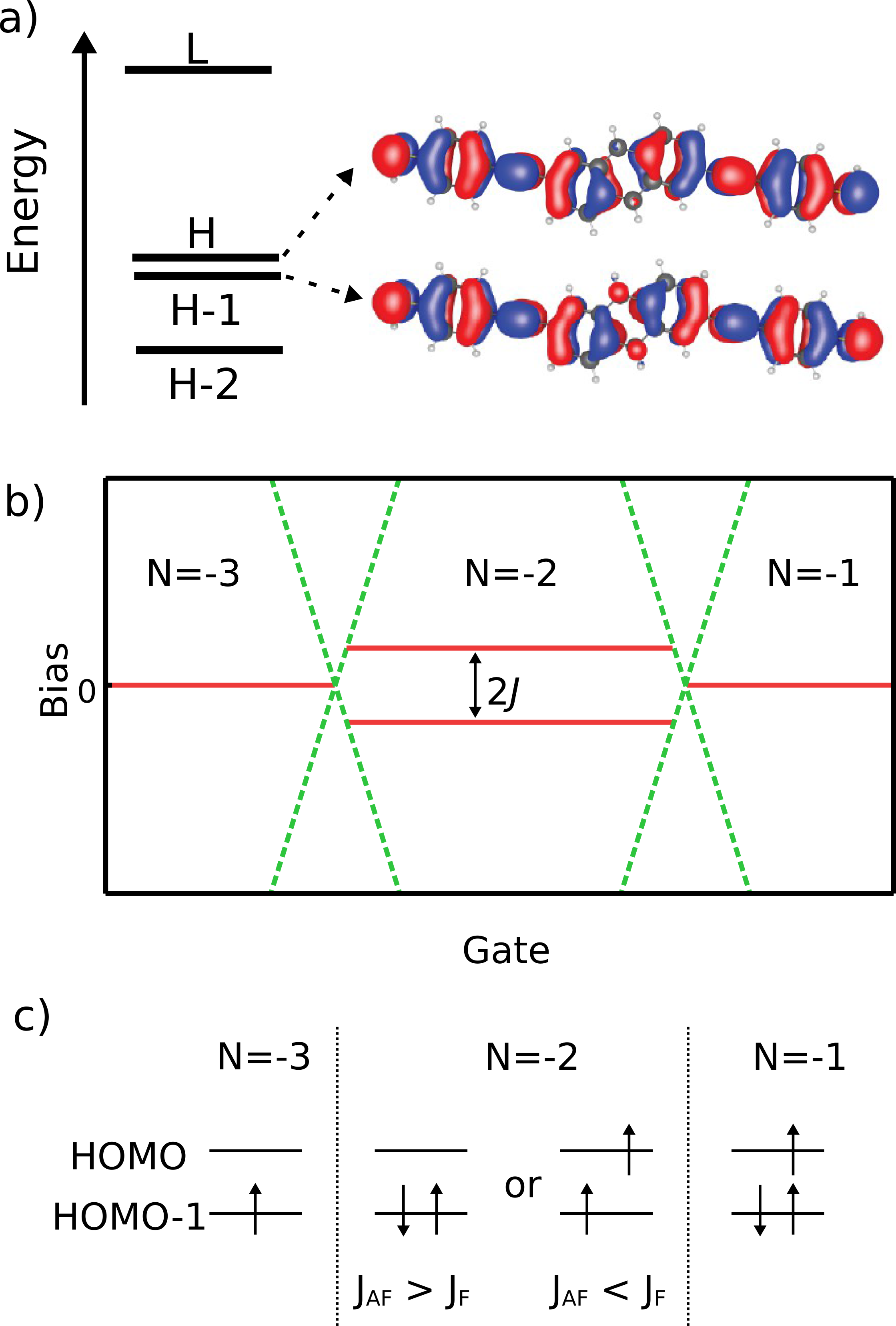}
	\caption{\label{figure:4} a) Energy spectrum of the AH molecular orbitals (not to scale); L stand for lowest unoccupied molecular orbital (LUMO) and H for highest occupied molecular orbital (HOMO). The isosurface of the nearly degenerate HOMO and HOMO-1 are shown next to it. It can be seen that the HOMO-1 is symmetric whereas the HOMO is anti-symmetric with respect to the molecule. b) Schematic of the transport features in AH at low temperature and $B = 0$~T. Green dashed lines are Coulomb edges; red lines represent higher-order transport features in the Coulomb blockade region, i.e, zero-bias peaks associated with $S=1/2$ Kondo or inelastic tunneling excitations at finite bias. c) Orbital filling of the nearly degenerate HOMO and HOMO-1. In the N = -2 charge state two ground states are possible depending on the exchange coupling.}
\end{figure}

The observation of low-bias features in a charge state with an even occupation suggests that the singlet to triplet energy difference for the two highest lying electrons is small. For molecules this may seem counter intuitive due to their small size, however, when inspecting the density functional calculations (DFT)\cite{Perrin2014} of the orbitals of AH (figure~\ref{figure:4}a) it can be seen that the HOMO and HOMO-1 are nearly degenerate in energy. Analogous to the prototypical hydrogen molecule, these nearly degenerate orbitals are formed by the hybridization of both conjugated halves of AH and thus show symmetric and anti-symmetric orbital wave-functions. The level splitting ($\Delta E$) of the HOMO and HOMO-1 is of the order of 10 to 50 meV, depending on the geometry of the molecule (see the supplementary information of \citet{Perrin2014}), which is significantly smaller then in fully conjugated molecules. The level splitting gives an estimate for the internal coupling ($\tau$) between the two conjugated halves of the molecule by using $\tau = \Delta E/2$. This results in the coupling of the conjugated moieties to be between 5 to 25 meV.  
 
We attribute the features in the measurements to the sequential oxidation of AH from a N = -1 occupation down to N = -3 occupation (where N is the difference in electrons with the neutrally charged molecule). Oxidation of the AH is supported by DFT calculations which predict the Fermi energy to lie close to the HOMO orbital. Furthermore, multiple reduction of AH is unlikely as the LUMO and LUMO+1 are not close in energy. This leads us to the following features in transport (figure~\ref{figure:4}b) and the electronic states (figure~\ref{figure:4}c) as a function of electron occupation of the HOMO and HOMO-1 in AH. In the N = -3 state, a single unpaired electron is present in the HOMO-1, which results in spin-1/2 Kondo in transport. Adding an electron results in the formation of a singlet or triplet state. Depending on the ground state, the excitation visible in the N = -2 charge state is the triplet or singlet respectively (this will be discussed further on). Addition of a second electron results in a pair of electrons in the HOMO-1 and a free unpaired spin in the HOMO; spin-1/2 Kondo reappears. No samples were measured that showed the three consecutive charge states shown in figure~\ref{figure:4}b, however, all eight samples show the N = -2 charge state. Next to the N = -2 charge state samples S1, S3, S7 and S8 show the N = -3 state and samples S2, S5 and S6 show the N = -1 charge state. 

The fact that both the singlet and triplet ground states are observed at $B = 0$~T, indicates that the ferromagnetic $J_F$ and anti-ferromagnetic $J_{AF}$ terms contribute almost equally to the total exchange coupling $J = J_F - J_{AF}$. To first order\cite{Launey2014}, $J_F$ is governed by the Coulomb exchange and $J_{AF}$ by the kinetic exchange. The kinetic exchange can be approximated by the Hubbard model, for $t < U$, $J_{AF} = 4t^2/U$, where $U$ is the on-site Coulomb interaction and $t$ is the inter-site hopping. Using the addition energy as an approximation for the on-site Coulomb interaction $U \approx \Delta E = 40-80$~meV and using $t\approx\tau = 10$~meV as calculated from DFT, one finds $J_{AF} = 5-10$~meV. A singlet ground state can thus be explained by considering only this term. However, the appearance of the triplet ground state suggests that the Coulomb exchange should also be taken into account. This can be intuitively understood from the fact that AH is approximately 2 nm long, much smaller than for other double quantum dot implementations. Therefore, electrons are closer together thereby increasing Coulomb effects and promoting ferromagnetic coupling (Hund's rule).

Further indications for a nearly degenerate HOMO and HOMO-1 can be seen in the off-resonant transport in the N = -1 and N = -3 charge states of S2 and S7 respectively. Here, in addition to the spin-1/2 Kondo resonance, low-bias excitations are visible. These can be related to the excitation of an electron from the HOMO-1 to the HOMO\cite{Begemann2010}. In both samples the excitation energy changes as a function of gate voltage; a possible mechanism for this could be differential gating\cite{Osorio2010} of the two conjugated halves of AH that form the HOMO and HOMO-1. If AH is situated asymmetrically in between the electrodes, the left conjugated half of AH can exhibit a slightly different gate coupling than the right conjugated half. It would thus modify the energy splitting ($\Delta E$) between the HOMO and HOMO-1, leading to low-lying gate dependent excitations as observed in the experiment.    

The delocalization and splitting of the HOMO and HOMO-1 and the observation of Kondo effects, suggest that the measurements show a DQD that has a strong internal coupling ($\tau$) and is well coupled to the leads ($\Gamma$). To further probe the spin states in AH and study spin relaxation mechanisms, it is beneficial to significantly lower $\Gamma$ and $\tau$ so that the spins on the two halves are localized more. This has been done in semiconductor\cite{Ono2002} and carbon nanotube\cite{Churchill2009} DQD's, where with Pauli spin blockade the spin dynamics and relaxation mechanisms have been studied. This localization can be achieved by using chemical synthesis to decrease the coupling to the leads\cite{Danilov2008} or the intramolecular coupling of the molecular DQD by using for example a larger non-conjugated bridge\cite{Mujica-Martinez2013} or a rotation of the two halves\cite{Venkataraman2006}.

In conclusion, we have shown transport spectroscopy of a strongly coupled double quantum dot molecule formed in a symmetric molecule with a non-conjugated bridge in the middle. A pattern of consecutive charge states has been identified which shows spin-1/2 Kondo, singlet/triplet states, spin-1/2 Kondo as a function of increasing electron occupation. This pattern can be explained with the formation of a pair of nearly degenerate orbitals due to the broken conjugation. In the even charge state transport features are observed, which indicate the presence of singlet and triplet states. The ground state varies from sample to sample, indicating that the anti-ferromagnetic and ferromagnetic interactions are of the same order of magnitude in AH. This suggest that Coulomb exchange plays a significant role as can be expected from a double quantum dot the size of a single molecule.

\begin{acknowledgments}
This research was performed with financial support from the Netherlands Organization for Scientific Research (NWO/OCW), FOM and by an ERC advanced grant (Mols@Mols).
\end{acknowledgments}

\end{document}